\documentclass{article}

\usepackage[english]{babel}

\usepackage[a4paper,top=2cm,bottom=2cm,left=3cm,right=3cm,marginparwidth=1.75cm]{geometry}

\usepackage{amsmath}
\usepackage{graphicx}
\usepackage[colorlinks=true, allcolors=blue]{hyperref}

\title{From Bjorken Scaling to Scaling Violations}
\author{Giorgio Parisi}

\usepackage{graphicx}
\newcommand {\bi} {\bibitem}
\newcommand {\be} {$$}
\newcommand {\ee} {$$}
\newcommand{\beq}{\begin{equation}} \newcommand{\eeq}{\end{equation}}

\newcommand {\eps} {\epsilon}

\newcommand {\lan} {\langle}
\newcommand {\ran} {\rangle}

\newcommand {\bc} {\begin{center}}
\newcommand {\ec} {\end{center}}
\newcommand {\bd}{\begin{displaymath}}
\newcommand {\ed}{\end{displaymath}}

\def \form#1 {eq. (\ref{#1}) }
\def \parziale#1#2 {{\partial {#1} \over \partial {#2}}}
\def \bi#1 {\typeout{#1} \item}

\begin{document}
\date{}
\maketitle
\begin{center}
    Dipartimento di Fisica,  Università degli Studi di Roma {\sl La Sapienza}\\
Istituto Nazionale di Fisica Nucleare, Sezione di Roma I, \\
Institute of Nanotechnology (NANOTEC) - CNR, Rome unit
\end{center}
\vspace{.5cm}
\begin{abstract}
    This paper traces the historical and conceptual journey from Bjorken scaling to the discovery of scaling violations in deep inelastic scattering, culminating in the development of Quantum Chromodynamics (QCD). Beginning with the challenges faced by early strong interaction theories in the 1950s, we explore the emergence of agnostic approaches such as the bootstrap philosophy and current algebra, which sought to describe hadronic phenomena without relying on specific field theories. The pivotal role of experimental results from SLAC in the late 1960s is highlighted, leading to Bjorken's proposal of scaling in deep inelastic scattering and Feynman's parton model. We then delve into the theoretical breakthroughs of the 1970s, including Wilson's operator product expansion and the renormalization group, which provided the framework for understanding scaling violations. The discovery of asymptotic freedom in non-Abelian gauge theories by Gross, Wilczek, and Politzer marked a turning point, establishing QCD as the theory of strong interactions. Finally, we discuss the formulation of the Altarelli-Parisi equations, which elegantly describe the evolution of parton distribution functions and scaling violations, and their profound impact on the study of hard processes in particle physics. This paper not only recounts the key developments but also reflects on the interplay between theory and experiment that drove the field forward.
\end{abstract}

\section{Introduction}
It is a well-known fact that at the beginning of '50 people were losing hope of a microscopic theory of strong interactions.
In 1953, after spending nearly one year with his students on some computation in perturbative pion nucleon theory Dyson went to Fermi to ask an opinion \cite{dyson2004meeting}; Fermi said:

{\sl With the pseudoscalar meson theory there is no physical picture, and the forces are so strong that nothing converges. To reach your calculated results, you had to introduce arbitrary cut-off procedures that are not based either on solid physics or solid mathematics.}

{\it In desperation I asked Fermi whether he was not impressed by the agreement between our calculated numbers and his measured numbers. He replied, “How many arbitrary parameters did you use for your calculations?” I thought for a moment about our cut-off procedures and said, “Four.” He said, “I remember my friend Johnny von Neumann used to say, with four parameters I can fit an elephant, and with five I can make him wiggle his trunk."}

The reason was clear as stressed by Fermi: the coupling constant was so large and one could not compute more than the first and second order in the perturbative expansion. The discovery of many resonances (e.g. the $\Delta$) made clear that the situation was far from a perturbative framework. Moreover, nothing indicated that the nucleon and the pion were more fundamental than other particles.

There was no convincing suggestion for having a particular type of field theory. Moreover, if a wizard could tell which was the basic field theory, such knowledge would be considered useless as far as there was no way to exploit it: no reliable techniques existed beyond perturbation theory.
As remarked by David Gross {\sl until 1973 it was not thought proper to use field theories without apologies} \cite{gross2005nobel}.  

This paper aims to recall the various steps that led to the resurrection of Lagrangian field theories.

\subsection{ Agnostic theories}: 

. 

 In the absence of a reliable theory, people started to study the consequences of a generic field theory without committing to a particular one. Many different keywords were used: symmetry groups, axiomatic field theory dispersion relations, S-matrix, Regge poles, superconvergence sum rules, and bootstrap, but there was no space for field theories with Lagrangians, with well-defined equations of motion.

 In one case, there was a well-defined mathematical setup where one could deduce rigorous results. The basic idea was that some fields act as a creation operator of a given particle; these fields may not be fundamental they could be composite. In other words, for a particle of type $\alpha$ there is an interpolating field $\psi_\alpha(x)$ such that
 \begin{equation}
     \langle 0|\psi_\alpha(x)|\alpha\rangle \ne 0 \,.
 \end{equation}
The fields are local, i.e. the commutator (or the anticommutators) are zero at space-like distances, as it should be in a Lagrangian field theory.

This approach was very successful. 
\begin{itemize}
    \item One proved that the CTP had to be exact \cite{r1964pct}.
    \item The Lehmann-Symanzik and Zimmermann (LSZ) formalism \cite{lehmann1957formulation} could be used to extract the S matrix from the vacuum expectation of the product of the interpolating field.
    \item Mostly surprising in some kinematical range it was possible to prove the validity of the dispersion relation that relates the imaginary and real part of the scattering amplitudes \cite{mandelstam1958determination}.
\end{itemize}

Around 1960, Geoffrey Chew introduced the {\sl bootstrap philosophy}, which posited that there were no elementary constituents; all particles were considered equally fundamental.   
  According to bootstrap theory, every particle is {\sl composed} of all other particles; there is a {\sl democracy} among elementary particles, and none is more fundamental than any other.  The age-old search for the constituent elements of matter had come to an end; there were no longer constituent elements of matter, only relationships between the various particles. It was an idea that proved to be very successful.  

A book on the bootstrap approach even suggested that a detailed knowledge of quantum field theory could hinder the adoption of new ideas. This stance was somewhat paradoxical, given that dispersion relations—a cornerstone of the bootstrap—were originally derived within the context of local field theory. Ironically, the bootstrap approach played a pivotal role in the emergence of string theories, which are among the most sophisticated quantum field theories.

 Murray Gell-Mann himself was not taking quarks seriously: he was considering them more as a mathematical model to implement $SU(3)$ symmetry. In 1964, he wrote \cite{gell1964symmetry} {\sl We use the method of abstraction from a Lagrangian field theory
model. In other words, we construct a mathematical theory of
the strongly interacting particles, which may or may not have
anything to do with reality, find suitable algebraic relations that
hold in the model, postulate their validity, and then throw away
the model. We compare this process to a method sometimes
employed in French cuisine: {\it a piece of pheasant meat is cooked
between two slices of veal, which are then discarded}.}

The absence of elementary point-like objects suggested that hadrons were extremely soft. This viewpoint was confirmed by the very fast decay of the proton form factor and by the exponential suppression of particle productions at large momentum transfer. Hagedorn theory \cite{hagedorn1965statistical} predicted a maximum temperature somewhat less than 200 MeV: also a very energetic hadronic fireball would emit hadrons of no more than a few hundredth MeV. The exponentially arising spectrum was later explained as the effect of the transition that leads to the creation of a quark-gluon plasma \cite{cabibbo1975exponential}.

\subsection{There was also a different viewpoint}.
 

Electromagnetism, Fermi interactions, and the V-A theories for weak interaction were based on local currents and quantum field theory. The purely leptonic electro-weak decays were described by a 4-Fermions field theory that was non-renormalizable; however, there was some hope that the introduction of heavy vector Bosons could make the theory renormalizable.

A hadronic current mediated the semi-leptonic weak interaction, and the resulting current algebra (based on local commutators) was crucial to normalizing the weak interaction vertices, which played a fundamental role in 1963 Cabibbo's theory of weak interaction \cite{cabibbo1963unitary}; indeed the normalization of the weak current was possible only after the identification of the current inside the generators of the symmetry group $SU(3)$. An angle $\theta$ among the currents was previously suggested in 1960 by Gelmann and Levy \cite{gell1960axial}: unfortunately, they have not introduced a symmetry group and concluded that {\sl There is, of course, a renormalization factor for that decay, so we cannot 
be sure that the low rate really fits in with such a picture.} The group $SU(3)$ was introduced in 1961: the identification of the weak vector current with the current that generates the $SU(3)$ group was impossible at the time of \cite{gell1960axial}.

The quantum field theory approach to physics was strongly pushed in Europe: there were many collaborations among different scientific institutions that were later formalized in the Triangular Meetings (Paris-Rome-Utrecht).

\section{Crucial Steps to Bjorken Scaling}
The path to the formulation of Bjorken scaling involved numerous pivotal developments. A key aspect was understanding the joint properties of the electromagnetic and weak currents, as well as exploring the consequences of their local commutators, particularly concerning axial current normalization.

This period saw a decade of intense research. Below, I highlight some of the most significant contributions:
\begin{itemize}
    \item (1958) The discovery of the V-A theory for $\Delta S=0$ semileptonic transitions (Feynman, Gell-Mann) \cite{feynman1958theory}. The currents in the $\Delta S=0$ sector resembled those of the leptonic sector.
    \item (1963) Cabibbo's theory \cite{cabibbo1963unitary} extended this framework to $\Delta S=1$ semileptonic transitions, identifying the relevant currents in this sector.
    \item (1964) Ademollo and Gatto \cite{ademollo1964nonrenormalization} proved that, at first order in $SU(3)$ symmetry breaking, the matrix elements of the vector currents remain unrenormalized. This result validated a key approximation in Cabibbo's theory.
    \item (1964) Gell-Mann \cite{gell1964symmetry} introduced the $SU(3)\times SU(3)$ symmetry, generalizing the $SU(2)\times SU(2)$ symmetry proposed by Güraly and Radicati \cite{gursey1964spin}.
    \item (1965) Dashen and Gell-Mann \cite{dashen1965approximate} emphasized the importance of local current commutators, leading to the development of current algebra. While current algebra arises naturally in Lagrangian theories, no realistic Lagrangian framework existed at the time.
    \item (1965) In a seminal paper, Fubini, Furlan, and Rossetti \cite{fubini1965dispersion} demonstrated that local commutators imply sum rules. These sum rules simplify in the infinite momentum frame, a concept introduced in this work. The infinite momentum frame later played a crucial role in developments such as the parton model.
    \item (1966) Bjorken \cite{bjorken1966applications} gave a comprehensive analysis of the implications of current algebra. The following year, he published a review paper \cite{bjorken1967current} with the highly suggestive title: {\sl Current Algebra at Small Distances}.
    \item (1968) Callan and Gross derived a sum rule for the cross-section in deep inelastic scattering \cite{callan1968crucial}, as a direct consequence of current commutators. This formula was later generalized by Cornwall and Norton \cite{cornwall1969current}, who derived additional sum rules.
    \item (1969) Bjorken proposed the concept of {\sl Bjorken scaling} for deep inelastic scattering, building on earlier sum rules \cite{bjorken1969asymptotic}. This proposal will be discussed in detail in the following section.
    \item (1969) In a groundbreaking paper, Feynman introduced the parton model \cite{feynman1969very}. Technically, Feynman's approach mirrored field theory, positing the existence of fundamental objects. However, he was making the implicit (and impossible) assumption that the interactions were super-normalizable, eliminating divergences. The framework was formulated in the infinite momentum frame, making it more tractable. The fundamental fields were unknown, but some consequences could still be explored. This approach provided a simplified framework, free from the complexities of renormalization and logarithmic divergences that Feynman himself had introduced in electrodynamics.
    \item (1969) Shortly after Feynman's proposal, Bjorken and Paschos outlined the implications of partons in a paper titled {\sl Deep Inelastic Scattering in the Parton Model} \cite{bjorken1969inelastic}. The parton model emerged as the simplest explanation for deep inelastic scaling and the associated sum rules, which could be straightforwardly derived in the large momentum limit.
\end{itemize}

\section{Deep Inelastic scattering}
 The acceleration of the theoretical results at the end of the sixties was driven by the experimental results on deep inelastic scattering \cite{breidenbach1969observed}. In 1966 the Electron Linear Accelerator of SLAC  started to work. In 1969 the electron beam reached an energy of 17 GeV. 
  The process studied was $electron+nucleon \to electron+hadrons$. At the leading order in quantum electrodynamics, it is equivalent to $virtual Photon+nucleon \to hadrons$.
 
 In the interesting region, the process is highly inelastic. The wonderful idea was to look only at the energy and momentum of the final hadrons without studying all other observables. However, one needed to study the energy and momentum loss of the scattered electron, so no hadron detector was necessary. There was only one electron detector at a fixed angle, that  could be changed.

The only two relevant parameters were the mass squared of the virtual photon  ($-q^2$) and its energy loss $\Delta E$. It was convenient to introduce the parameter $\nu\equiv \Delta E M$, $M$ being the nucleon mass. One was interested in the region where both parameters are large, away from elastic scattering and from real photon scattering. It is convenient to write:
$x=q^2/\nu$. The variable $x$ is kinematically constrained to belong to the interval $0-1$.

At the end, one can write the cross-section in terms of two functions of these two variables $F_1(q^2,x)$, $F_2(q^2,x)$. Looking the the dependence on the angle one could separate the two contributions.

\subsection{Bjorken scaling}

Let us concentrate on the function $F_2(q^2,x)$, also because it gives the most relevant contribution.

Roughly speaking, Bjorken assumed that some equal-time commutators of currents and their derivatives have a non-zero vanishing element. After some computations, he found that
\begin{equation}
\lim_{q^2\to\infty} \int _0^1 dxF_2(q^2,x) x^{n-1} =M_{n}\,,
\end{equation}
(the case $n=1$ is related to standard equal-time commutators of currents).

The quantities $M_n$ are equal to the matrix element of the commutators. If one makes the simplest and innocent-looking hypothesis that matrix elements of the commutators  are non-zero, one arrives at the formula
\begin{equation}
 \lim_{q^2\to\infty} F_2(q^2,x)= F_2^\infty (x)\,, \qquad \int _0^1 dxF_2^\infty(x) x^{n-1} =M_{n}\,.
\end{equation}

Finally, Bjorken and Paschos showed the function $F_2^\infty (x)$ was proportional to the fraction of charged partons carrying a fraction $x$ of the momentum of the proton in the $P=\infty $ frame, the factor of proportionality being the squared charge of the parton. The parton properties and distribution in the infinite momentum frame were observed in a conceptually simple experiment.

 The asymptotic independence of $F_2(q^2,x)$  on $q^2$ was in reasonable agreement with the data. The experimentalists liked very much the proposal: of course, there was some dependence on $q^2$, however, this was small and it was supposed to disappear in the asymptotic large momenta region.

\section{ Subtle is field theory}

\subsection{Lacking of scaling in perturbative field theory}

The main problem to fix was the identification of the partons and their interaction. A quite natural proposal was the identification of partons with quarks, but other identifications were also possible. The troubling point was the choice of the interaction. Indeed the partons are very similar to free particles, or to particles interacting with a superrenormalizable interaction that could be neglected at high energy: in renormalizable theories, Bjorken scaling was not perturbatively correct: strong logarithmic (and maybe power) corrections were present. Unfortunately in 4 dimensions, no super-renormalizable theories with Fermions are available. The approach recalls what happened during the discovery of quantum mechanics \cite{parisi2001planck}; bold (and wrong) hypotheses were first made and the progress made using the hypothesis led to the correct conclusions.

In the sixties, beyond perturbation theory, the large momentum behavior of the Green functions was a mystery. The renormalization group was relevant but the consequences of the renormalization approach were not understood.
 In a famous book on electrodynamics \cite{bjorken1965relativistic} Bjorken and Drell wrote that in electrodynamics if the function $\beta(e^2)$ has a zero at $e_0^2$, the bare charge is given by
\begin{equation}
    e_0^2=F^{-1}(\infty) \label{fixed}
\end{equation}
 {\sl We are presented with a dilemma. All our arguments in this section have been based on the renormalization program in perturbation theory, which permits us to expand propagators, vertex functions, etc, in power series in both the normalized and bare charges $e$ and $e_0$. However, in the previous equation, we have come up with a result that puts a condition on the value of $e_0$ indicating that it cannot chosen arbitrarily. The behavior is completely foreign to the perturbation development and forces us to conclude that at least one of our assumptions along the way has been wrong. (...) There, conclusions based on the renormalization group arguments concerning the behavior of the theory summed at all orders are dangerous and must be viewed with due caution. So is it with all conclusions from local relativistic field theories. } 

 Nowadays equation (\ref{fixed}) seems perfectly logical. For Bjorken and Drell the theory was defined in terms of its perturbative expansion, while nowadays the theory is defined by the path integral formulation and the non-convergent \cite{itzykson1977asymptotic,t1979can} (but asymptotic) perturbative expansion is just a tool that is useful in the week coupling regime. 

\subsection{Operator dimensions and non-canonical scaling}

More or less at the same time Bjorken and Drell were writing their disconsolate conclusion (that echoed the same conclusion of Gell-Mann and Low), Ken Wilson had a completely different approach. His viewpoint was that strong interacting field theory exists and that one can compute its properties with due ingenuity. This viewpoint was clearly exposed in a magnificent and forgotten paper \cite{wilson1965model} of 1965, where many of the ideas of the later most famous papers are already present. 

In the introduction, he writes: {\sl The Hamiltonian formulation of quantum mechanics has been essentially abandoned in investigations of the interactions of mesons, nucleons,
and strange particles. This is a pity. (...) 
There are two reasons why the Hamiltonian approach
was discarded in the study of strong interactions. One
reason was that no one knew what Hamiltonian to use,
or how to obtain the correct Hamiltonian. The other
reason was the problem of renormalization: the problem
that whenever one tried to solve a Hamiltonian for a
Lorentz-invariant theory, particle self-energies and the
like were infinite.
}

In this paper, he discusses the solution of a field theory model introducing a sequence of scale-dependent Hamiltonians. The renormalization group is seen as the functional relation of the Hamiltonian at one scale with the one at another scale. In this constructive approach, the Gell-Mann Low Bjorken Drell paradox fades out. The bare coupling constant is the fixed point of this transformation: nothing strange that its value is fixed.

\subsection{Wilson's operator expansion}
More or less simultaneously (1964) a revolutionary idea enters in the game: the so-called Wilson operator expansion.
The idea is written in a one-hundred-page preprint with the title {\sl On Products of Quantum Field Operators at Short Distances} \cite{wilson1964products} that appeared as a Cornell Report and it was never published \footnote{It was submitted to Phys. Rev.: there was a long referee report to which Wilson started to write the reply, but he never finished it.}.
The paper was not a success. It was noticed by Zimmerman and it was at the basis of Brandt's PhD thesis. Indeed it was first cited by Brandt in 1967 \cite{brandt1967derivation}, the second citation was in 1970: now it has around 40 citations. 

Wilson's short-distance operator product expansion states that in the limit of small $x$ the product of two operators can be written as
\begin{equation}
A(x) B(0) \to_{x\to 0} \sum_C C(0) |x|^{-d_A-d_B+d_C} \,.
\end{equation}
The leading terms come from the operators $C$ with the lowest dimensions.
The dimensions of the operators were the canonical ones in free theory where everything was clear. The bold hypothesis was that such an expansion is valid also in strong coupling field theories, where the operator dimensions may be different from the canonical ones. \footnote{Similar ideas were rediscovered in the context of phase transition by Migdal and Polyakov: operator fusion.}

Wilson's ideas became popular much later (1969), with the magnificent paper:
{\sl Non-Lagrangian models of current algebra} that has more than 3000 citations and was an immediate success \cite{wilson1969non}. In the introduction, he writes:
{\it Field theories with exact scale invariance are not
physically interesting since they cannot have finite mass particles. But one can hypothesize that there exists
a scale-invariant theory which becomes the theory of
strong interactions when one adds mass terms to the
Lagrangian the strong interactions become scale-invariant at short distances. This leads to the idea of broken scale invariance proposed by Kastrup and Mack.}

Later on: {\sl
It is assumed that the strong interactions contain
some arbitrary fundamental parameters just as the mass
and charge of the electron are fundamental parameters
in electrodynamics. However, the greater complication
of strong interactions means that the parameters of
strong interactions are not physical masses and coupling constants; they show up explicitly only in the short-distance behavior of strong interactions. Implicitly, they
determine all of the strong interactions, but to calculate
physical masses and coupling constants one has to solve
the strong interactions, which is not possible at present.
In physics these parameters have particular values, but
the theory of strong interactions is assumed to be selfconsistent for any values of the parameters. In particular, if all the parameters are zero, it will be assumed that
all partial symmetries become exact. The theory with
all free parameters set equal to zero will be called the
"skeleton theory".}

{\sl Skeleton theory}, a forgotten word, radically changed the perspective: all masses are zero, and the phenomenology of resonances disappears. In skeleton theory, the fundamental objects are the {\sl off-shell} Green functions.
 Wilson is speaking of fields (e.g. the pion field), and their dimensions. 
 
 At the end of the paper, he discusses the vertex of the axial current with the two electromagnetic currents (the vertex is crucial for the determination of the amplitude of the process $\pi_0\to \gamma \gamma$). {\sl It is hard to imagine that one could have a complete
formula for this vertex function without having a complete solution of the hadron skeleton theory. The prospects for obtaining such a solution seem dim at present}.

So the physically motivated renormalization group (that had tremendous success in the field of second-order phase transitions) and the Wilson operator expansion set the stage for a non-perturbative understanding.

New interest in the renormalization group arose in 1970 with the Callan-Symanzik equation \cite{callan1970broken,symanzik1970small}. 
The Callan-Symanzik was more explicit than the renormalization group and much easier to understand. It seemed less paradoxical than the renormalization group. It worked very well with the massive theory.

In the 1971 paper, Symanzik \cite{symanzik1971small} proved in one particular case the Wilson operator expansion for the scalar $\phi^4$ theory:

\begin{equation}
\phi(x)\phi(0) \approx C(x) \phi(0)^2\qquad x\to 0\,.
\end{equation}
where the behavior of the function $C(x)$ was controlled by an anomalous exponent $\gamma_2(g)$.

The argument was that in a Green function when the two momenta go to infinity and the others remain fixed, the large momentum behavior had anomalous terms that can be controlled analytically. If we translate these results into configuration space we obtain the previous equation. In conclusion in 1971 the Wilson expansion was taken for granted and there were no doubts about its validity.

\subsection{ The light cone expansion}
 In 1971  Brant and Preparata discovered  \cite{brandt1971operator} that Bjorken scaling was related to the so-called {\sl Light cone expansion}.
The first observation was kinematical:
 The value of the cross-section in deep inelastic scattering is related to configuration space to the singular behavior near the light cone $x^2=0$ of the function
\begin{equation}
\lan p| J(x) J(0) |p\ran \,,
\end{equation}
where $J$ is the electromagnetic current (vector indices are neglected) and $|p\rangle$ is the one proton state.  Using  the Wilson expansion near $x=0$, the behavior near the light cone could be obtained by  looking at many terms:
\begin{equation}
    J(x)J(0)\to_{x^2\to 0}=\sum_{a=0,\infty} O^a{\mu_1, \dots, \mu_a} x^{\mu_1}\dots x^{\mu_a} C_a(x)\,.
\end{equation}
It is possible to show that each term in the infinite sum over $a$ corresponds to a different moment of the experimental structure-function.

Bjorken scaling follows if the lightcone singularities are the same as in free theory, i.e. if the operators that enter into the Wilson expansion have canonical dimensions. The requirement that the equal time commutators of the derivatives of the current have non-zero matrix elements is replaced by the requirement that the operator in the Wilson expansion have canonical dimensions. So the technical hypothesis on the commutators becomes a global hypothesis on the operator dimensions of the theory.

In this way, the Wilson short-distance operator product becomes deeply related to the experimentally observed approximate Bjorken scaling.

 In the case of canonical dimensions, one obtains a simple form for the light cone expansion:
\begin{equation}
J(x)J(0)\to_{x^2\to 0} \frac {O(x,0)} {x^2}\,,
\end{equation}
where $O(x,0)$ is a bilocal operator.
In free field theory  a simple form of the bilocal operator is found. For example in a scalar theory:
\begin{equation}
    O(x,0)=\varphi(x) \varphi(0) \,.
\end{equation}

\subsection{ What happens in an interacting theory?}

Christ-Hasslacher-Mueller  (1972) computed \cite{christ1972light} the coupling-dependent anomalous dimensions of the operators relevant for deep inelastic scattering (the so-called twist-two operators) for a pseudoscalar theory and for a theory with a simple vector interaction.

If we neglect the dependence of the running coupling constant on the momenta, one has something like
\begin{equation}
M_n(q^2)\equiv\int _0^1dxx^{n-1}F(x,q^2) \ ; \ \ \ M_n(q^2)=C_n \exp ( \gamma_n (\alpha) \log(q^2)) \,.\end{equation}

The dependence on the running coupling constant can be trivially added. If one assumes that in the large momentum region, the running coupling constant $\alpha$ goes to a fixed point value $\alpha_c$, one gets:
\begin{equation}
     M_n(q^2)=C_n (q^2)^{\gamma_n (\alpha_c)} 
\end{equation}
Canonical scaling implies that $\gamma_n (\alpha_c)=0$.
So Bjorken Scaling could be obtained if $\gamma_n (\alpha_c)=0$.

\section{The quest for a scaling invariant theory}

\subsection{A scaling invariant theory with canonical dimensions is free}
This part of the paper is more based on personal recollections, they were strongly influenced by the physicists whose papers I was reading and I was in contact with, and in particular by Ken Wilson's ideas.

If strong interactions at large momenta are scaling invariant and the operators have well-defined operator dimensions, a natural question was: can a scaling invariant non-free theory have $\gamma_n (\alpha_c)=0$ for all $n$?

The answer (although not under the form of a theorem) was no \cite{ferrara1972canonical,parisi1972serious,parisi1972bjorken,parisi1973measure}.
So only two possibilities were present: 
\begin{itemize}
    \item The scaling invariant skeleton theory is free and we have Bjorken scaling.
    \item The scaling invariant theory skeleton is not free and we  do not have Bjorken scaling
\end{itemize}

Symanzik \cite{symanzik1973field} noticed that the $\lambda \phi^4$ with a negative coupling constant is an example of a theory that has an asymptotically computable behavior (asymptotically free in modern language): the running coupling constant being at the leading order $\lambda(q^2)=\lambda/(1-C \lambda \log(q^2/m^2)$, with positive $C$, so that in the large $q^2$ region $\lambda(q^2)\approx 1/(C \log(q^2/m^2))$.
In this theory, Bjorken scaling is asymptotically exact without logs \cite{parisi1973computable}.

The theory is not realistic: it is unstable and no Fermions are present. Theories of Fermions interacting with scalar or pseudoscalar particles were not asymptotically free, as remarked by Landau in the 50's. So a strong interacting theory was missing.

\subsection{The illusion of a scaling invariant theory for strong interaction}

In non-asymptotically free theory the short distance theory should correspond to a fixed point of the renormalization group.
If this happens, one would be very far from the perturbative region and there is no clear distinction between the fundamental field and the composite fields.  For some time there was the hope that the skeleton theory could be found by solving some kind of consistency equations, some kind of bootstrap equations of a novel kind, using all the symmetry of the problem, and in particular conformal invariance \cite{ferrara1972covariant} \footnote{The approach worked for 2-dimensional scalar theory and much later for 3-dimensional scalar theory, but at that time there was not the capacity to do these computations.}. 

What about quarks? The quark field would have a non-zero Green function and the quark field acting on the vacuum would produce a state with a non-zero quark number. At that time quark confinement was not known, so the absence of quark production was interpreted as a high mass for the quarks: hadrons should be deep bounded states of quarks. In the scaling invariant skeleton theory bringing these quarks to zero mass would be a too-violent approximation (a more reasonable approximation could be to bring their mass to infinity). Indeed in the 1968  Wilson's paper, in the list of fields, no quark field was present. All the fields listed corresponded to physical particles, currents, and the stress tensor.
So the strong interacting skeleton theory was naturally quarkless.

In such a theory Bjorken scaling is not correct. However, a careful analysis showed that strong violations of Bjorken scaling were compatible with experiments. \cite{parisi1973experimental}. Indeed most of the good experimental data were in the region where both $q^2$ and the mass of the final hadrons were large. Indeed large $q^2$ and a large produced mass were needed to stay in the asymptotic region: given the experimental limitation that implied an $x$ neither too large nor too small. 
Some operators must have canonical dimensions also in strong interacting skeleton theory: the current and the stress tensors (this leads to two sum rules of the kind that we have previously discussed). One finds that the $F(x,q^2)$ should decrease near $x=1$ and increase at small $x$. Obviously, in the central $x$ the function should be nearly constant. This was known from 1973, where it was also remarked that strong violations of Bjorken scaling were possible, and this field theory motivated approach {\sl  may
be an explanation of the fact that the "scaling limits"
seem to be reached from above at large $x$, and from
below at small $x$} \cite{parisi1973experimental}. In that same paper, it was noticed that, if only one operator for given quantum numbers contributes, one could write equations of the form
\be \frac{\partial F(x,q^2)}{ \partial \log(q^2)}=\int_x^1 \frac{dy}{ y} F(y,q^2) K(x/y)\,,\ee
The kernel $K$ depends on the value of the anomalous dimensions of the operators.

This was a simple rewriting of the formula for the moments. It is useful for phenomenological analysis because the value of the function at a small $x$ does not enter the formula (in the moment's formulation, we have to integrate over all range of $x$). However, its simple meaning in terms of effective parton distribution was not contained in that paper. The idea of evolving parton densities appeared (as far as I know) firstly \cite{polyakov1971similarity} and in \cite{kogut1974parton}.

\subsection{Asymptotic freedom was discovered}

All these dreams of a strong interacting theory faded away when the negative sign of the beta function of the renormalization group was discovered, and a viable model of strong interaction theories with asymptotic freedom was proposed.

The negative sign for the beta function was discovered independently three times.
\begin{itemize}
\item In 1969 Iosif Khriplovich presented a beautiful computation that could be done on the back of the envelope \cite{khriplovich1969green}. He computed the vacuum energy $E$ induced by a constant gauge field. $E$ is trivially $F^2$ at the tree level. At one loop level, he computed the shift in the vacuum energy by studying the quantization of free particles in the field (Landau levels). The formulae were well known and one had to take care that gluons have spin one. At the end, he got
\begin{equation}
    E=F^2(1-A\alpha \log(q^2/\mu^2))\,.
\end{equation}
From this formula, the value of the beta function could be directly obtained. The computation was done with on-shell particles, so there was no need to use the Fadeev-Popov ghosts.
\item  At a conference in Marseilles in the summer of 1972,  Gerard ’t Hooft, announced that he had calculated the sign of the beta function in Yang-Mills and that the result was negative!  This grand announcement was met with indifference. There were few people present, and even they did not pay much attention. 
	The only person really capable of understanding ’t Hooft’s result was  Symanzik, who urged him to write an article on the subject.   ’t Hooft had begun work on some extremely difficult calculations on quantum gravity.  For him, the beta function calculation was hardly more than an exercise, and he did not have time to write it up for publication.

\item In 1973, as everybody knows, by Gross and Wilczek \cite{gross1973ultraviolet,politzer1973reliable}, and Politzer.
\end{itemize}
Interestingly, the results on the negative sign of the coupling theories were noticed by most scientists in the reverse order of their discovery.
However, only the last papers make the connection with colored gluons and show that asymptotic freedom is valid for Quantum Chromodynamics. Indeed, only these authors did the computation, aiming to apply it to strong interaction theory.

Let me add a personal touch to the story \cite{parisi2023flight}. I happened to be good friends with Symanzik.  In November 1972 I went to visit him for two weeks in Hamburg.  Surprisingly, he did not talk to me about ’t Hooft’s result. Symanzik wanted the result to be communicated to the world directly by ’t Hooft in a written form.
It was only in February 1973 that I learned from Symanzik about ’t Hooft’s result.   I had just moved to CERN in Geneva for two months, and since ’t Hooft was there, we met.

	We needed to identify the gauge theory of strong interactions and verify that the beta function was negative.  It sounded easy; in 1972 \cite{fritzsch1973advantages} it was proposed that quarks existed in three different ‘colors’ interacting by exchanging colored gluons: essentially the Yang-Mills theory studied by ’t Hooft.  I knew this theory perfectly, but the arguments were based on the naive parton model, a free skeleton theory.  I had put my money on the opposite hypothesis – a non-free skeleton theory – and very presumptuously I had put down this proposal as too naïve.  Then I shelved it.
    
	Looking back on it now, the conversation with ’t Hooft was surreal. We discussed only the possibility that the gauge theory was the flavor $SU(3)$. This was nonsense and we immediately realized it. After half an hour of discussion, we concluded that we could not construct a model for the strong interactions were the negative sign of the beta function.
	We did not give a moment’s thought to the colored gluons.   We were incredibly blind, for which I take full responsibility, because I knew by heart the experimental work and the various models proposed in the literature.  Unfortunately, I was not fast enough to realize that if the skeleton theory was free, all the conclusions coming from the assumption of a non-free skeleton theory should be canceled.

\section{The final expression for Scaling Violations}
\subsection{The first formulae for scaling violations.}
In 1973 Georgi and Politzer \cite{georgi1974electroproduction}, Gross, and Wilczek \cite{gross1974asymptotically} computed the Christ-Hasslacher-Mueller formulae for QCD.  One gets a simple result in the case of the non-singlet contribution: taking care of the running coupling constant $\alpha(q^2)$, the result is:
 
 \begin{equation}
     \frac{\partial M_n(q^2)}{ \partial \log(q^2)}= \gamma_n (\alpha(q^2))\, ,
 \end{equation}
 
 Using the same chain of arguments of my previous paper,  I finally got \cite{parisi1974detailed}: 
\be{\partial F(x,q^2)\over \partial \log(q^2)}=\int_x^1 {dy\over y} F(x,q^2) K(x/y,\alpha(q^2))\,,\ee
where kernel  $K(z,\alpha(q^2))$ (that was not interpreted as the quark fragmentation function) is the inverse Mellin transform of the anomalous dimensions $\gamma_n (\alpha(q^2))$.
Using the first order in $\alpha$ for $\gamma_n (\alpha)$, one finally gets for the non-singlet contribution:
\be
K(z,\alpha(q^2))=\frac83\frac{\alpha(q^2)} {4 \pi}\left( {1+z^2\over(1-z)_+} +\frac32 \delta (z-1)\right)\ee

\begin{figure}
    \centering
    \includegraphics[width=0.5\linewidth]{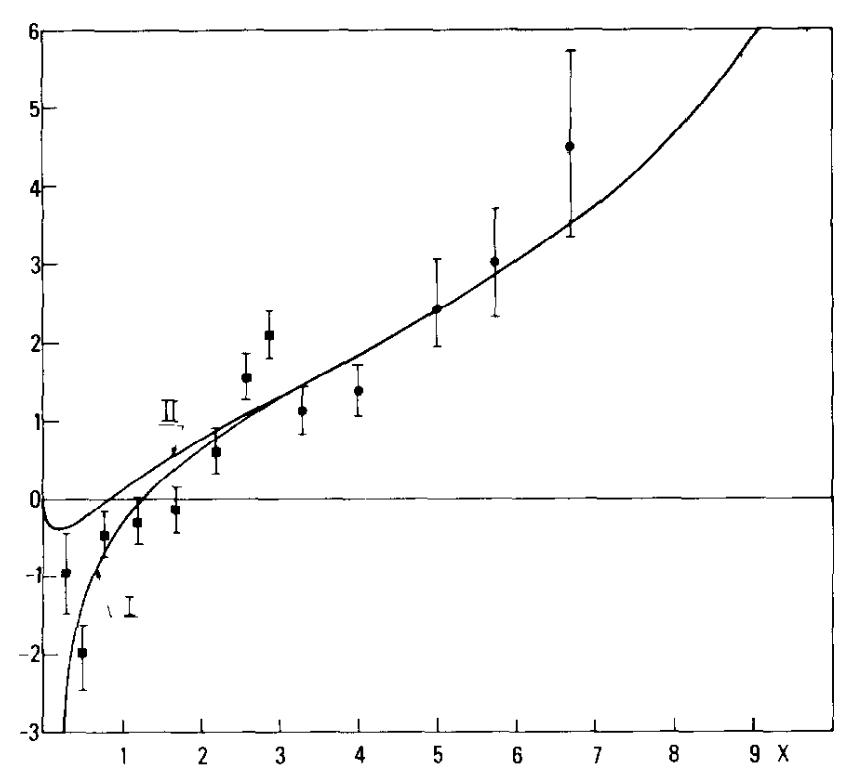}
    \caption{
    The prediction of Parisi and Petronzio \cite{parisi1976breaking} compared with the experimental data of SLAC \cite{riordan1974tests} for the plot
of $d \ln (F (x,q^2))/d \ln(q^2 )$
for the proton versus $x$. Curve I is their prediction for the logarithmic derivative of the proton structure
function compared with experimental data. Curve II is obtained by retaining only the octet operators in the operator
expansion. The difference comes from the contribution of the gluons, which is obtained by an educated guess.
    \label{fig:enter-label}}
\end{figure}

 The extraction of the value of $\alpha$ using the formulae for the moments 
was done in 1976 \cite{parisi1976breaking}. Using the moment formulation and the Mellin transform, we computed the violations of scaling without any physical interpretation of the formulae (Parisi Petronzio 1976) and we got  $\alpha_s=0.4$. Curve I of fig. \ref{fig:enter-label} was our prediction for  $d \log(F_2^p(x,q^2)/d \log(q^2)$ compared to the experimental data. Curve II was obtained by retaining only the octet operator in the operator expansion.  This was written before Altarelli-Parisi.

 \subsection {The Altarelli Parisi equations }

The problem of logarithmic corrections in the perturbative expansion was quite old. There was a different case where the physical interpretation was quite clear: quantum electrodynamics, where electromagnetic radiative corrections are small but not negligible. In QED, one often finds that radiative corrections are proportional to $\alpha \log(E/m_e)$ or $(\alpha \log(E/m_e))^k$. Various techniques can be used.
 The most famous approach is the {\sl Equivalent photon approximation in quantum electrodynamics} (due to Weizsaker-Williams.).

Nicola Cabibbo was strongly interested in computing these effects for applications to the experimental analysis of events collected in $e^+e^-$ collisions.
Cabibbo and Rocca in 1974 wrote \cite{cabibbo1974rho} formulae like:

\begin{equation} {P_{e\to e\gamma}(\eta)}=\frac{\alpha}{\pi}\frac{1+(1-\eta)^2}{\eta} \log( E/m_e)\, ; \ \ \ P_{\gamma\to e^+ e^-}(\eps)=\frac{\alpha}{2\pi}(1+(1-2\eps)^2) \log( E/m_e)\,,
\end{equation}
where $\eta$ is the fraction of longitudinal momentum carried by the photon
and $\eps$ is the fraction of longitudinal momentum carried by the electron.

 These probabilities can be combined. They computed the probability of finding inside a $\gamma$ a triplet $\gamma, e^+,e^-$: it is proportional to $P_{\gamma\to e^+ e^-} P_{e\to e\gamma}\log( E/m_e)^2$. Nicola and I 
 used these ideas (unpublished) to estimate the cross section for the process $ e^+e^- \to 2e^+2e^-$ that was a candidate to be an important background in $e^+e^-$ colliding beam experiments. 

Influenced by this paper, I extended the formulae for valence quarks to the gluons and sea-quarks \cite{parisi1976Moriond,parisi1992field}. In this case, one introduces effective quark and gluon parton distributions $N(x,\log(q^2))$:

\begin{eqnarray}
    \frac{dN_{q_i}(x,\log(q^2)}{d\log(q^2)}=\frac{\alpha}{4\pi}\int_x^1\frac{dy}{y}\left[ p_{qq}(x/y)N_{q_i}(x,\log(q^2))+ p_{qg}(x/y)N_{g}(x,\log(q^2))\right]  \,,
    \nonumber\\
    \frac{dN_{g}(x,L)}{d\log(q^2)}=\frac{\alpha}{4\pi}\int_x^1\frac{dy}{y}\left[ p_{gg}(x/y)N_{g}(x,\log(q^2))+ p_{qg}(x/y)\sum_{i}N_{q_i}(x,\log(q^2))\right]\,.\label{AP}
\end{eqnarray}

The equations are formally the same as the Altarelli-Parisi (apart from a few typos), however, the derivation was completely different.  The proof was formulated in the language of renormalization group equations for the coefficient functions of the local operators which appear in {\sl the light cone expansion for
the product of two currents}. Therefore, the equations were useless for other hard processes. Most of the relations coming from the parton model were out of reach of the operator product expansion.

 The physical derivation of the formulae using the ideas of effective parton distribution appeared in my paper with Guido; this paper stemmed from a proposal from Guido, i.e. to make previously-obtained results on scale violations clearer and more exploitable. 
 
The motivations of the paper were clearly stated in the introduction.
{\sl In this paper, we show that an alternative derivation of all results of current interest for the $Q^2$ behavior of deep inelastic structure functions is possible. In this approach {all stages of the calculation refer to parton concepts} and offer a very illuminating physical interpretation of the scaling violations. (\dots)
This method can be described as an appropriate generalization of the equivalent photon approximation in quantum electrodynamics (Weizsaker-Williams .... Cabibbo-Rocca).}
 
 {Indeed 
 in the spring '77} Guido suggested that it would be pedagogically useful to derive all the equations for scaling violations using the same techniques of Cabibbo-Rocca; { no loops were involved: only the evaluation of the vertices in the infinite momentum frame.}  Most of the formulae we needed were written there: only the gluon splitting into two gluons function was missing.
 
 The final master \footnote{We used the wording {\em master equations} because they were classical probabilistic equations derived in a quantum setting.} equations were essentially those of eq. \ref{AP}. The computations were particularly transparent and simple when we extended them to the case of polarized partons using the helicity formulation in the infinite momentum frame.

The paper was a well-done cocktail of renormalization group results, parton model, and perturbation theory in an infinite momentum frame: easy to drink and to swallow.  It was really pedagogic: it contained all the logical steps and the final receipt was quite easy to follow \cite{forte2025asymptotic}.

\subsection{Aftermath}

I am convinced that the most important result of the paper was {\sl not} the construction of a practical way to compute scaling violations in deep inelastic scattering.  The crucial point was to shift the focus from Wilson operator expansion to the effective number of partons that was dependent on the resolution, i.e. $q^2$. {It was more than a computation: it was a {\sl change in the language} we use. The appropriate choice of language is one of the most important scientific tools.

For example, the Drell-Yan process (i.e. $pp \to \mu^+ \mu^-+\cdots$) could be not studied by a Wilson operator expansion: the Brandt-Preparata analysis did not work in this case. Similar conclusions are also true for the jet production in hadronic collisions and for the computation of the cross-section for large transverse momenta that were related to the hard scattering of partons.

However, it was now possible to study these processes using the new language: we had to factorize the amplitude for the process into a part containing the effective parton distribution at the relevant energy and into a part containing the hard scattering that could be treated in perturbation theory in the running coupling constant. Moreover, in the presence of energy-dependent effective parton distributions, the parton model predictions were ambiguous at order $\alpha_{QCD}$ and next to the leading order corrections (NLO) have to be computed to fix the predictions.

 The solution to all these problems was at hand after our paper. This opportunity was immediately taken by Guido. All the following papers are published in '78 just after our '77 paper:
\begin{itemize}
\item{\it Leptoproduction and Drell-Yan processes beyond the leading approximation in chromodynamics} \cite{altarelli1978leptoproduction}. 

\item{\it Transverse momentum of jets in electroproduction from quantum chromodynamics} \cite{altarelli1978transverse}.

\item{\it Transverse momentum in Drell-Yan processes} \cite{altarelli1978Drelltransverse}.

\item{\it Processes involving fragmentation functions beyond the leading order in QCD} \cite{altarelli1979processes}. 
\end{itemize}

In all these four problems, the matching of the experimental data with the theoretical predictions allowed an independent measure of the QCD running coupling constant. At that time, it was possible to achieve this goal in the first three cases. The compatibility of these independent determinations of $\alpha_{QCD}$ with the value coming from deep inelastic scattering (and from $\psi$ decay) was instrumental in convincing physicists that QCD was the correct theory.

\bibliography{main}
\bibliographystyle{ieeetr}
\end{document}